\documentclass[a4paper,12pt]{article}
\usepackage[cp1251]{inputenc}
\usepackage[T2A]{fontenc}
\usepackage[english,russian]{babel}
\usepackage{amsmath,amsfonts}
\usepackage{amsthm}
\usepackage{mathtext}
\usepackage[psamsfonts]{amssymb}
\usepackage{bm}
\usepackage{indentfirst}
\usepackage{geometry}
\usepackage{graphicx}
\usepackage{cite}
\usepackage{enumerate}
\usepackage{units}

\renewcommand{\d}{\textrm{d}}
\newcommand{\const}{\textrm{const}}

\newtheorem*{Corollary*}{Следствие}

\begin{document}
{\selectlanguage{english}

UDC 532.11; 539.19
\begin{center}
{\Large {\bfseries Pressure Operator for the P\"{o}eschl--Teller Oscillator}}

\bigskip
{\large Yu. G. Rudoy, E. O. Oladimeji}
	
\medskip
Department of Theoretical Physics and Mechanics\\
Peoples' Friendship University of Russia\\
6, Miklukho-Maklaya str., Moscow, 117198, Russia

\end{center}

\noindent
\emph{Abstract}:
The quantum-mechanical properties of the strongly non-linear quantum oscillator in the P\"{o}eschl--Teller model are considered. In the first place, the energy spectrum and its dependence upon the confinement parameter (i.e., the width of the ``box'') are studied. Moreover, on the grounds of the Hellman--Feynman theorem the pressure operator in this model is obtained and (along with the energy spectrum) is studied in two main approximations: the ``particle in the box'' and ``linear (harmonic) oscillator'' for large and low values of the main quantum  number; the critical value is also evaluated. Semi-classical approximation as well as perturbation theory for the P\"{o}eschl--Teller are also considered. The results obtained here are intended for future thermodynamic calculations: first of all, for the generalization of the well-known Bloch result for the linear harmonic oscillator in the thermostat. To this end, the density matrix for the P\"{o}eschl--Teller oscillator will be calculated and the full Carnot cycle conducted.

\medskip
\noindent
\emph{Keywords}: Bloch and P\"{o}eschl--Teller quantum oscillator, pressure operator, Hellman--Feynman theorem, quasi-classical approximation, harmonic oscillator, particle in a box.


\section{Formation of the Model}\label{sec:1}

Some decades ago P\"{o}eschl and Teller~\cite{key-1} introduced a family of anharmonic PT-potentials $\mho(x)$, which allowed the exact solutions of the one-dimensional Schr\"{o}dinger equation in the coordinate, or $x$-representation~\cite{key-2,key-3}. One of the most interesting member of the PT-family is the symmetric trigonometric potential, which is even in the variable $x$, where $-L<x<L$.
\begin{equation}\label{eq:1}
\mho(x,L)=\mho(-x,L)=\mho_{0}\tg^{2}[\alpha(L)x],
\end{equation}
\begin{equation*}
\alpha(L)=\pi/2L.
\end{equation*}

At $x=\pm L$ the potential becomes singular, which physically means the presence of a pair of impenetrable walls. They confine the movement of the non-relativistic particle with positive constant mass m. The parameters $\mho_{0}$ and $L$ are also positive, though the limits $\mho_{0}\rightarrow 0$ and $L\rightarrow \infty$ are also allowable and will be considered properly.

The presence of the walls is of specific interest for the future thermodynamic description of this model, placed into some thermostat. In contrast to ordinary harmonic oscillator (HO) for the PT-oscillator (PTO) it is possible to introduce the pressure operator $\hat{P}(\hat{x},\hat{p},L)$ which according to Hellmann and Feynman~\cite{key-4,key-5} is connected with the energy operator or the Hamiltonian $\hat{H}(\hat{x},\hat{p},L)=(\nicefrac{\hat{P}^{2}}{2m})+\mho(x,L)$ by the formal relation:
\begin{equation}\label{eq:2}
\hat{P}(\hat{x},\hat{p},L)=-\frac{\partial}{\partial L}\hat{H}(\hat{x},\hat{p},L).
\end{equation}

Strictly speaking, one should differentiate in~\eqref{eq:2} with the volume $\mho=L^{d}$, where $d$ is dimension of the coordinate space, where $d=1$ will be held here everywhere. The important point is that the operation $(\nicefrac{\partial}{\partial L})$ may be fulfilled only after the operator $\hat{H}$ has acted upon some wave function $\phi$ --- as a rule, upon the eigenfunction $\phi_{n}(x)$ of $\hat{H}$ with eigenvalue $E_{n}$ (see, e.g., the detailed analysis in~\cite{key-4,key-5}). The importance of this sequence of operations is directly connected with the account of boundary conditions $\psi_{n}(\pm L)=0$ at all values of $n$. In this case the formal definition~\eqref{eq:2} acquires more definite sense:
\begin{equation}\label{eq:3}
P_{n}(\hat{L})=-\frac{\partial E_{n}}{\partial L},
\end{equation}
if $\hat{H}\psi_{n}(x,L)=E_{n}\psi_{n}(L)$.

Note, that in this paper we won't be engaged with the eigenfunctions $\psi_{n}(\pm L)$ --- it is sufficient to know that all of them contain the factor $\cos\alpha x$, which ensures the fulfillment of zero boundary condition at the walls $x=\pm\nicefrac{\pi}{2\alpha}=\pm L$. Following~\eqref{eq:7} one may see, in particular, that if the energy spectrum $E_{n}(L)$ is a uniform function of $L$ (e.g. $E_{n}(L)\thicksim L^{-s}>0$) so that $P_{n}(L)=(\nicefrac{s}{L})E_{n}(L)$, or in the operator form:
\begin{equation}\label{eq:4}
\hat{P}(L)=s\hat{h}(L),
\end{equation}
where $\hat{h}(L)=\nicefrac{\hat{H}(L)}{L}$ is the (linear) density of the energy; the last relation is the well known barocaloric equation of state for the ideal gas. Note that by obtaining (4) we have used Euler's uniformity property $(\nicefrac{\d}{\d L})\alpha^{m}(L)=(\nicefrac{m}{L})\alpha^{m}(L)$ valid for any real value of $m$.

\section{Exact Energy and Pressure Spectra}\label{sec:2}

It is remarkable that the complication potential~\eqref{eq:2}, leads to an exact solution of the Schr\"{o}dinger equation with potential~\eqref{eq:1} with fully discrete positive energy levels $E_{n}(L)>0$ $\bigl($including the ground levels $E_{1}(L)\bigr)$:
\begin{equation}\label{eq:5}
E_{n}^{PT}(L)=E_{n}^\text{FP}(L)+E_{n}^\text{HO}(L), \quad (n=1,2,3,\dots),
\end{equation}
obviously, the spectrum $E_{n}(L)$ is unbounded from above.

Two terms in~\eqref{eq:5} look like the free particle in the box and harmonic oscillator respectively:
\begin{equation}\label{eq:6}
E_{n}^\text{FP}(L)=T(L)n^{2}, \quad E_{n}^\text{HO}(L)=\hbar\omega(L)\biggl(n-\frac{1}{2}\biggr),
\end{equation}
note, that~\eqref{eq:6} doesn't contain terms of higher order in $n$ that quadratic one.

The quantity $T(L)$ is equal to the well known minimal kinetic energy of the $FP$ in the box with dimensions $\left[-L,L\right]:$
\begin{equation}\label{eq:7}
T(L)=E_{1}^\text{FP}(L)=\frac{\hbar^{2}}{2m}\alpha^{2}(L), \quad \alpha(L)=\frac{\pi}{2L}.
\end{equation}

The $\omega(L)$ quantity is the frequency of some HO, and depends upon $\alpha(L)$ in much more complicated way than~\eqref{eq:7}).

In this paper we won't be engaged with the eigenfunctions $\varphi_{n}(x)$ --- it is sufficient to know that all of them contain the factor $\cos\alpha x$, which ensures the fulfillment of zero boundary condition at the walls $x=\pm\frac{\pi}{2\alpha}=\pm L$.

Complicated method:
\begin{equation}\label{eq:8}
\hbar\omega(L)=T(L)\lambda(L),\lambda(L)=\biggl[\biggl(\frac{2}{\pi\zeta(L)}\biggr)^{2}+1\biggr]^{\nicefrac{1}{2}}-1,
\end{equation}
\begin{equation}\label{eq:9}
\zeta^{2}(L)=\frac{1}{\pi^{2}}\frac{T(L)}{\mho_{0}}=\biggl(\frac{\hbar}{\pi}\biggr)^{2}\frac{1}{2m\mho_{0}}\alpha^{2}(L).
\end{equation}

Obviously, that the formal structure of the parameter $\lambda(L)$ resemble the kinetic energy of the free relativistic particle with rest mass in ($(\zeta(L))$ is the like $p$, where $p$ is the particle's momentum) $\lambda(L)\geq 0$ $\lambda(L)=0$ only at the point $\frac{1}{\zeta(L)}=\frac{1}{\alpha(L)}=0$ in the limit $L\rightarrow \infty$).

Following the definition~\eqref{eq:3}, one finds from~\eqref{eq:5} the exact diagonal matrix elements of the pressure operator~\eqref{eq:2}:
\begin{equation}\label{eq:10}
P_{n}^{PT}(L)=P_{n}^\text{FP}(L)+P_{n}^\text{HO}(L), \quad P_{n}^\text{FP}(L)=\frac{2}{L}E_{n}^\text{FP}(L),
\end{equation}
\begin{equation}\label{eq:11}
P_{n}^\text{HO}(L)=\frac{2}{L}E_{n}^\text{HO}(L)-\frac{1}{L}T(L)\psi(L)\biggl(n-\frac{1}{2}\biggr),
\end{equation}
the Euler's uniformity property:
\begin{equation}\label{eq:12}
\psi(L)=[\lambda(L)+1]\left\{ 1-[\lambda(L)+1]^{-2}\right\} \geq 0,
\end{equation}
where we have used Euler's $(\nicefrac{\d}{\d L})\alpha^{m}(L)=(\nicefrac{m}{L})\alpha^{m}(L)$

The expressions $(E^\text{HO})$ and $(P^\text{HO})$are rather complicated since $\lambda(L)$ is \textit{non-uniform} function of $L$ pressure operator $\widehat{P}(L)$ for $PTO$ is in general not proportional to the energy operator $\widehat{H}(L)$, but in extreme case ($FP$ and $H$-parts) this property is restored.

At fixed $L$, the relative contribution of the $FP$ and HO depends upon $n$ and it's determined by the ratio:
\begin{equation}\label{eq:13}
\eta_{n}(L)=\frac{E_{n}^\text{FP}(L)}{E_{n}^\text{HO}(L)}=\frac{n}{n_{cr}(L)}, \quad n_{cr}(L)=\frac{1}{\lambda(L)}.
\end{equation}

Clearly, at $\eta_{n}(L)\ll 1$, $n\ll n_{cr}(L)$ i.e for lower energy levels, the HO- part dominates, whereas at $\eta_{cr}(L)\gg1,n\gg n_{cr}(L)$, i.e, for higher energy levels, the $FP$-part dominates. This is easy to understand, because at $V_{0}$ and $L$ held constant the growth of the particle's energy $E$ makes the potential~\eqref{eq:1} more and more the limiting potential $\mho(x,L)=\delta(x-L)+\delta(x+L)$, which characterizes the FP in the box.

In the last case we obtain he fully free particle without any ``box'', so the particle's energy is not quantized at all. The same limit at fixed $n$ is achieved at $V_{0}=0$.

Moreover, the $FP$-limit full at fixed $n$ is achieved also at the limiting point $\alpha(L)=0$ or $\nicefrac{1}{L}=0$, because in this case the potential~\eqref{eq:1} also is identically equal to zero. However, one should note that the limit of small, but finite $\alpha(L)\ll 1$ resembles more not the FP-, but the HO- case.

\section{FP- and HO-limits for the Energy and Pressure Spectra}\label{sec:3}

It is instructive to consider the expressions~\eqref{eq:5} and~\eqref{eq:10} at fixed values of $n$ in two basic limiting cases, i.e. FP in the box and HO-limits.

\subsection{FP in the box limit:}

$L=\const$, $V_{0}\rightarrow 0$, $\pi\zeta(L)\rightarrow \infty$. In this case, $T(L)$ is large compared to $V_{0}$, so $\nicefrac{1}{(\pi\zeta(L))}=V_{0}T(L)\ll 1$, thus:
\begin{equation}\label{eq:14}
\lambda(L) \approx  \frac{2}{\bigl(\pi\zeta(L)\bigr)^{2}}\left[1-\frac{1}{2}\frac{2}{\bigl(\pi\zeta(L)\bigr)^{2}}\right], \quad \lambda(L)\ll 1.
\end{equation}
Due to the general definition~\eqref{eq:8}, the effective frequency is of the form:
\begin{equation*}
\hbar\omega(L)\approx 2V_{0}\left[1-\nicefrac{2V_{0}^{2}}{T(L)}\right]
\end{equation*}
and tends to zero with $V_{0}$.

For the pressure operator $\hat{P}$ in this limit the whole term~\eqref{eq:11} is negligible, so for $\hat{P}^\text{FP}$ the linear operation, equation of state~\eqref{eq:4} holds, where $s=2$ and
\begin{equation*}
\hat{H}_{(x,p)}^\text{FP}=\frac{\hat{P}^{2}}{2m}+\left[\delta(x-L)+\delta(x+L)\right].
\end{equation*}

\subsection{HO-limit}

$V_{0}=\const$, $L\rightarrow \infty$, $\pi\zeta(L)\sim\alpha(L)\rightarrow 0$. In this case, $T(L)\sim\alpha^{2}(L)$ is small compared to the effective frequency in the lowest order HO- approximation $\hbar\omega(L)=T(L)\sim\lambda_{0}\alpha^{2}(L)$ here:
\begin{equation*}
\lambda(L)\approx\frac{2}{\pi\zeta(L)}\sim\frac{1}{\alpha(L)}
\end{equation*}
is a large quantity. More precisely, from~\eqref{eq:8} follows that
\begin{equation*}
\lambda(L)\approx\tilde{\lambda}(L)\biggl[1-\frac{1}{\tilde{\lambda}(L)}+\frac{1}{2}\frac{1}{\bigl(\tilde{\lambda}(L)\bigr)^{2}}\biggr], \quad \lambda(L)\ll 1,
\end{equation*}
further, using again~\eqref{eq:8}, one obtains
\begin{equation}\label{eq:15}
\hbar\omega(L)\approx\hbar\tilde{\omega}(L)-T(L)+\frac{1}{2}\frac{T(L)}{\hbar\omega(L)}.
\end{equation}

Here:
\begin{equation}\label{eq:16}
\hbar\tilde{\omega}(L)=T(L)\tilde{\lambda}(L)=2\left[V_{0}T(L)\right]^{\frac{1}{2}}=\alpha(L)\biggl[\frac{2V_{0}\hbar^{2}}{m}\biggr]^{\nicefrac{1}{2}}
\end{equation}
is of order $\alpha(L)$, while the second and third term in~\eqref{eq:15} are of the order $\alpha^{2}(L)$ and $\alpha^{3}(L)$ accordingly.

The pressure operator $\hat{P}$ in this limit may be found from~\eqref{eq:15} by noticing that both terms in the rhs of the equation make contributions of opposite sign, but of the same (lowest) order $\alpha(L)$. Indeed, in this limit $T(L)\psi(L)\approx T(L)\lambda(L)=\hbar\tilde{\omega}(L)$ and
\begin{equation}\label{eq:17}
E_{n}^\text{HO}(L)\approx\tilde{E_{n}}^\text{HO}(L)=\hbar\tilde{\omega}(L)\biggl(n-\frac{1}{2}\biggr).
\end{equation}

Combining relations~\eqref{eq:16} and~\eqref{eq:17}) one finds for the pressure operator $\tilde{P}^\text{HO}$, as well as for $\tilde{P}^\text{FP}$, the linear operator equation of state~\eqref{eq:4} holds, but now with $s=1$ and effective Hamiltonian:
\begin{equation*}
\tilde{H}_{(\hat{x},\hat{p})}^\text{HO}=\frac{\hat{P}^{2}}{2m}+\frac{1}{2}m\tilde{\omega}^{2}(L)x^{2},
\end{equation*}
which describes some ``confined'' HO.

\section{Approximations for the Energy Spectrum}\label{sec:4}

\subsection{Quasi-classical approximation (QC)}

It is instructive to compare the exact energy spectrum~\eqref{eq:5} with its QC-counterpart, for which the quantization rule states that for $n=1$, 2, 3, \dots.
\begin{equation}\label{eq:18}
\beth(E)=2\hspace{-8pt}\int\limits_{-x_{0}(E)}^{x_{0}(E)}\hspace{-8pt}p(x,E)\,\d x=2\pi\hbar\biggl(n-\frac{1}{2}\biggr),
\end{equation}
here $\beth(E)$ is the classical action for the PT oscillator with the potential energy~\eqref{eq:1} while $p(x,E)$ is the classical momentum.
\begin{equation}\label{eq:19}
p(x,E)=\sqrt{2m}\left[E-\mho_{0}\tg^{2}\alpha(L)x\right]^{\nicefrac{1}{2}}, \quad
p(x_{0},E)=0,
\end{equation}
obviously, $x_{0}(E)\rightarrow \pm L$ as $E\rightarrow \infty$.

The Bohr--Sommerfeld quantization takes the explicit form:
\begin{equation}\label{eq:20}
\biggl(\frac{1}{\alpha(L)}\biggr)\sqrt{2m}\left[(E+V_{0})^{\nicefrac{1}{2}}-V_{0}^{\nicefrac{1}{2}}\right] = \hbar\biggl(n-\frac{1}{2}\biggr),
\end{equation}
so that the QC energy spectrum will be of the same form as in~\eqref{eq:5}--\eqref{eq:6}. The only difference is that the exact quantity $\lambda(L)$ in the QC-case will be substituted by it's QC-analogy $\bigl[\frac{2}{\pi\zeta(L)}\bigr]+1$. It is easy to verify that, all the basic features of the PT-oscillator remains just the same (up to some numerical factors).

\subsection{Perturbation theory}

Consider the expansion of the potential:
\begin{equation}\label{eq:21}
\mho(x,L)=\mho_{0}=(\alpha(L)x^{2})\biggl[\biggl(1+\frac{2}{3}\alpha(L)x^{2}\biggr)\biggr]+\Bigl[(\alpha(L)x)^{4}\Bigr],
\end{equation}
which is plausible when both $\alpha(L)$ and $x$ are small. The leading term in~\eqref{eq:21} may be written down as the usual HO-potential:
\begin{equation*}
V_{0}\alpha^{2}(L)x^{2}=\frac{1}{2}m\tilde{\omega}^{2}(L)x^{2},
\end{equation*}
where the frequency $\tilde{\omega}^{2}(L)$ is the same as defined by~\eqref{eq:15}. It may seem rather unpleasant that $\tilde{\omega}^{2}(L)$ goes to zero with $\alpha(L)$ at $L\rightarrow \infty$, and so one may ask for the usual constant HO-frequency $\omega_{0}=\sqrt{\nicefrac{k}{m}}$ $(k=\const)$.

Such a result may be achieved simply by means of the potential intensity ${\mho_{0}\rightarrow (\frac{k}{2})\alpha^{-2}(L)}$. Note, that in this case as the limit point $\alpha(L)$. One arrives strictly to the usual HO-oscillator (but not the PT-oscillator). Indeed, the suggested rescaling can't save ``all the next order terms in the expansion''~\eqref{eq:21}. Moreover, at this limiting point the notion of pressure (the operator as well as it's spectrum becomes meaningless, so that we don't use this ``scaling trick'' --- as well as the point $\alpha(L)=0$ in the followings (though the HO-limit $\alpha(L)\ll 1$ is quite appropriate)). Consider now the term of the lowest order $(\alpha(L)x)^{4}$ in~\eqref{eq:21} as the weak an harmonics them the energy spectrum is of the form~\cite{key-2,key-3}:
\begin{equation}\label{eq:22}
E_{n}(L)=\tilde{E}_{n}^\text{HO}(L)+\Delta E_{n}(L), \quad \tilde{E}_{n}^\text{HO}(L)=\hbar\tilde{\omega}(L)\left(n+\frac{1}{2}\right),
\end{equation}
\begin{equation*}
\Delta E_{n}(L) = \biggl(\frac{3}{2}\biggr) \biggl[V_{0}\alpha^{2}(L)\biggr(\frac{2}{3}\biggr) \alpha^{2}(L)\biggr] \biggl(\frac{\hbar}{m\tilde{\omega}(L)}\biggr)^{2} \biggl[n^{2}+n+\frac{1}{2}\biggr],
\end{equation*}
where $n=0$, 1, \dots.

Making the shift $n\rightarrow (n-1)$:
\begin{equation}\label{eq:23}
E_{n}^\text{HO}(L)=\hbar\tilde{\omega}(L)\biggl(n-\frac{1}{2}\biggr), \quad
\Delta E_{n}(L)=T(L)\biggl[n^{2}+n+\frac{1}{2}\biggr].
\end{equation}

Two effects are evident once from~\eqref{eq:23}. Firstly, the linear in $n$ part of $\Delta E_{n}(L)$ brings for the non-perturbated spectrum $\tilde{E}_{n}^\text{HO}(L)$ equal to $[-T(L)]$, which agrees with the correction of the same order in~\eqref{eq:2} the full expansion. Secondly, the quadratic in n part reproduces the term $E_{n}^\text{FP}(L)$ from the exact energy spectrum~\eqref{eq:6}. Unfortunately, the next approximation (e.g. of orders $(\alpha x)^{6}$ and/or $(\alpha x)^{8}$ tend to spoil these nice results. In particular, they bring in $E_{n}(L)$ so, ``non-physical'' terms of orders $n^{3}$, $n^{4}$, etc. and also deform the term $T(L)n^{2}$, which should not be affected at all.

\section{Conclusion}\label{sec:5}

In this paper the quantum-mechanical properties of the strongly non-linear quantum oscillator in the P\"{o}eschl--Teller model were considered. In Sec.~\ref{sec:1} the formulation of the model was given and its relations with two well known models --- i.e., the free particle in the box and the harmonic oscillator were considered. Sec.~\ref{sec:3} was devoted to the analysis of the fully discrete spectrum of the model; also, using the Hellman--Feynman theorem the pressure operator was obtained and analyzed. namely, both the ``free particle in the box'' and ``harmonic oscillator'' limits were given the detailed investigation. Finally, in Sec.~\ref{sec:4} some approximations for the energy spectrum are considered --- namely, the quasi-classical one as well as the perturbation theory in the region near the harmonic one --- i.e., were the anharmonic terms are relatively small. All the results obtained here will be used for the thermodynamic calculations in the nearest future publication.

}

\newpage
УДК 532.11; 539.19

\begin{center}
{\Large {\bfseries Оператор давления для осциллятора Пёшля--Теллера}}

\bigskip
{\large Ю. Г. Рудой, Е. О. Оладимеджи}
	
\medskip
Кафедра теоретической физики и механики\\
Российский университет дружбы народов\\
Россия, 117198, Москва, ул. Миклухо-Маклая, 6

\end{center}

Рассмотрены квантово-механические свойства сильно нелинейного квантового осциллятора в модели Пёшля--Теллера. Изучен энергетический спектр модели и его зависимость от параметра конфайнмента, или эффективной ширины потенциала. На основе теоремы Гельмана--Фейнмана получен оператор давления для указанной модели, который вместе с энергетическим спектром изучены в двух основных приближениях: частицы в ящике и линейного гармонического осциллятора для больших и малых значений главного квантового числа $n$ соответственно; получено также значение критического значения $n_\text{кр}$. Рассмотрены также квазиклассическое приближение и теория возмущений для обоих предельных случаев. Полученные результаты предназначены для использования в последующих термодинамических приложениях --- прежде всего, обобщения хорошо известного результата Блоха для линейного гармонического осциллятора в термостате. С этой целью необходимо построить матрицу плотности для осциллятора Пёшля--Теллера для проведения полного цикла Карно.

\medskip
\emph{Ключевые слова}: квантовый осциллятор Блоха и Пёшля--Теллера, оператор давления, теорема Гельмана--Фейнмана, квазиклассическое приближение, гармонический осциллятор, частица в ящике.

\end{document}